\pdfoutput=1
\documentclass[12pt,prb,showpacs,preprintnumbers,amsmath,amssymb,superscriptaddress]{revtex4-1}

\usepackage[utf8]{inputenc}
\usepackage[english]{babel}
\usepackage{graphicx} 
\usepackage{transparent}
\usepackage{amsmath}
\usepackage{amssymb}
\usepackage{epstopdf}
\usepackage{amsfonts}
\usepackage{bm}
\usepackage{times}
\usepackage{mathtools}
\usepackage{hyperref}
\usepackage{tabularx}
\usepackage{hhline}
\usepackage[caption=false]{subfig}
\usepackage[export]{adjustbox}
\usepackage{sidecap}
\usepackage{relsize}
\usepackage{nopageno}

\usepackage{setspace}

\usepackage{color}
\usepackage[dvipsnames]{xcolor}
\graphicspath{{./figures/MF_GKBA_iTEBD/summary_figs/}}
\usepackage{amsmath,amsthm,amssymb}
\usepackage{mathrsfs}
\usepackage{lmodern}
\usepackage{parskip}
\usepackage[displaymath, mathlines]{lineno}
\usepackage{xcolor} 
\definecolor{chartreuse}{rgb}{0.5, 1.0, 0.0}
\definecolor{colq}{rgb}{1.0, 0.22, 0.0}
\definecolor{deepjunglegreen}{rgb}{0.0, 0.29, 0.29}
\usepackage{ulem}

\newcommand*{\citen}[1]{%
  \begingroup
    \romannumeral-`\x 
    \setcitestyle{numbers}%
    \cite{#1}%
  \endgroup   
}

\let\oldequation\equation
\let\oldendequation\endequation

\renewenvironment{equation}
  {\linenomathNonumbers\oldequation}
  {\oldendequation\endlinenomath}

\newcommand*{\TNS}{Ta$_2$NiSe$_5$}

\begin{document}

\title{Imaging the coherent propagation of collective modes in the excitonic insulator Ta$_2$NiSe$_5$ at room temperature }
\author{Hope M. Bretscher}
\thanks{These authors contributed equally}
\author{Paolo Andrich}\thanks{These authors contributed equally, Corresponding author}
\affiliation{Cavendish Laboratory, University of Cambridge, Cambridge CB3 0HE, United Kingdom}
\author{Yuta Murakami}
\affiliation{Department of Physics, Tokyo Institute of Technology, Meguro, Tokyo 152-8551, Japan}
\author{Denis Gole\v{z}}
\affiliation{Center for Computational Quantum Physics, Flatiron Institute, New York, New York 10010, USA}
\affiliation{Faculty of Mathematics and Physics, University of Ljubljana, Jadranska 19, SI-1000 Ljubljana, Slovenia}
\affiliation{Jožef Stefan Institute, Jamova 39, SI-1000, Ljubljana, Slovenia}
\author{Benjamin Remez}
\affiliation{Cavendish Laboratory, University of Cambridge, Cambridge CB3 0HE, United Kingdom}
\author{Prachi Telang}
\author{Anupam Singh}
\author{Luminita Harnagea}
\affiliation{Department of Physics, Indian Institute of Science Education and Research, Pune, Maharashtra 411008, India}
\author{Nigel R. Cooper}
\affiliation{Cavendish Laboratory, University of Cambridge, Cambridge CB3 0HE, United Kingdom}
\author{Andrew J. Millis}
\affiliation{Center for Computational Quantum Physics, Flatiron Institute, New York, New York 10010, USA}
\affiliation{Department of Physics, Columbia University, New York, New York 10027, USA}
\author{Philipp Werner}
\affiliation{Department of Physics, University of Fribourg,
Fribourg 1700, Switzerland}
\author{A. K. Sood}
\affiliation{Department of Physics, Indian Institute of Science, Bangalore, Karnataka 560012, India}
\author{Akshay Rao}\thanks{Corresponding author}
\affiliation{Cavendish Laboratory, University of Cambridge, Cambridge CB3 0HE, United Kingdom}
\date{\today}

\begin{abstract}

\end{abstract}

{
\let\clearpage\relax
\maketitle
}


\setlength\parskip{1em}
\setlength\parindent{0pt}

\textbf{Excitonic insulators host a condensate of electron-hole pairs at equilibrium, giving rise to collective many-body effects. Although several materials have emerged as excitonic insulator candidates, evidence of long-range coherence is lacking and the origin of the ordered phase in these systems remains controversial. Here, using ultrafast pump-probe microscopy, we investigate the possible excitonic insulator Ta$_2$NiSe$_5$. Below 328 K, we observe the anomalous micrometer-scale propagation of coherent modes at velocities of $\sim$10$^5$~m/s, which we attribute to the hybridization between phonon modes and the phase mode of the condensate. We develop a theoretical framework to support this explanation and propose that electronic interactions provide a significant contribution to the ordered phase in Ta$_2$NiSe$_5$. These results allow us to understand how the condensate’s collective modes transport energy and interact with other degrees of freedom. Our study provides a unique paradigm for the investigation and manipulation of these properties in strongly correlated materials.}

\section{Introduction}

The excitonic insulator (EI) phase is a state of matter that was first proposed more than 50 years ago  [\citen{jerome1967excitonic,halperin1968possible,keldysh1968collective}]. It was theorized that below a critical temperature (T$_c$), weakly screened Coulomb interactions in small bandgap semiconductors and semimetals could lead to the formation of an exciton condensate in the ground state (Fig.~\ref{fig1}A). In the proposed theoretical framework, condensation would result from the spontaneous breaking of a U(1) continuous symmetry, which describes the separate conservation of charges in the valence and conduction bands. Therefore, this macroscopic state would be accompanied by characteristic collective excitations, including an amplitude (Higgs) mode and a gapless phase (Goldstone) mode, as is the case for other correlated systems, like superconductors.  The emergence of the latter collective mode would lead to intriguing quantum phenomena such as superfluidic transport [\citen{coleman2015introduction,Hanamura1974}].

A number of small bandgap or semimetallic transition metal chalcogenide (TMC) compounds have recently attracted interest as promising excitonic insulator candidates [\citen{cercellier2007evidence,monney2011exciton,Seki2014PRB,mor2017ultrafast,werdehausen2018coherent,kogar2017signatures,wakisaka2009excitonic,kaneko2013orthorhombic,lu2017zero}]. The reduced dimensionality of TMCs results in weak screening of the Coulomb interaction and consequently large exciton binding energies, which could lead to condensation at non-cryogenic temperatures. 
Among this family of materials, particular attention has been devoted to Ta$_2$NiSe$_5$ following hints of the potential existence of an EI phase below 328 K. [\citen{Seki2014PRB,mor2017ultrafast,werdehausen2018coherent, wakisaka2009excitonic,kaneko2013orthorhombic,lu2017zero,mor2018inhibition,chen2020doping,suzuki2020detecting,tang2020non,kim2020direct,volkov2020critical,saha2101photo,pal2020destabilizing}]. These works have provided insights into the properties in \TNS{}, for instance, by using equilibrium [\citen{Seki2014PRB, wakisaka2009excitonic}] and time-resolved [\citen{mor2017ultrafast,okazaki2018photo,tang2020non,suzuki2020detecting,saha2101photo}] angle-resolved photoemission spectroscopy (ARPES), by analyzing the effect of physical and chemical pressure [\citen{lu2017zero,pal2020destabilizing}], doping [\citen{chen2020doping}], and by detecting anomalies in charge transport [\citen{lu2017zero}], optical signatures [\citen{ning2020signatures,andrich2020ultrafast}] and phonon properties [\citen{werdehausen2018coherent,mor2018inhibition,kim2020direct,volkov2020critical,pal2020destabilizing,kim2020observation}] as a function of temperature.

However, despite numerous experimental and theoretical studies, the predominant driving mechanism of symmetry breaking in \TNS{}, and thus the existence and character of the EI phase itself, remains controversial. In real condensed matter systems, contrary to the idealized picture presented above, the Hamiltonian's continuous symmetry may be further reduced to a discrete symmetry due to electron-phonon interactions or a hybridization between the conduction and valence bands~[\citen{mazza2019nature,watson2019band}]. Additionally, the observed structural phase transition can be driven both by electronic interactions and by electron-phonon coupling~[\citen{kim2020direct,kim2020observation,andrich2020ultrafast,saha2101photo,ning2020signatures,volkov2020critical}]. All these interactions are in principle present in a real system, and quantitative differences in their relative magnitudes have a significant, qualitative impact on the properties of the collective modes in the ordered phase. For instance, the presence of additional symmetry breaking terms in the Hamiltonian forces a gap at $k\simeq$ 0 in the dispersion of the phase mode, suppressing the proposed supertransport properties [\citen{zenker2014fate}]. Theory suggests that if the transition is largely due to electron-phonon interactions, one would even expect the low-energy, acoustic-like phase mode to be largely suppressed, and an excitonic insulator phase would not be supported by the material [\citen{baldini2020spontaneous,watson2019band}]. These different pictures could be untangled through the detection of the condensate’s collective modes and the analysis of their transport properties and real-space dynamics. If the phase mode remains gapless or minimally gapped, then the continuous symmetry must remain largely conserved and the transition is predominantly driven by electronic interactions, meaning the system is close to the idealised EI phase. 

In this work, we use ultrafast, spatially-resolved pump-probe microscopy to study the real space dynamics of Ta$_2$NiSe$_5$. We show that photoinduced excitations in \TNS{} propagate coherently over distances up to 1 $\mu$m and with electron-like velocities, which we attribute to collective modes of the excitonic condensate and their hybridization with the lattice degrees of freedom. These results provide novel insights towards understanding the nature of the ordered phase in \TNS{}, potentially establishing a pathway for its manipulation.

\section{Results}
   
\begin{figure*}[tbh]

\includegraphics[]{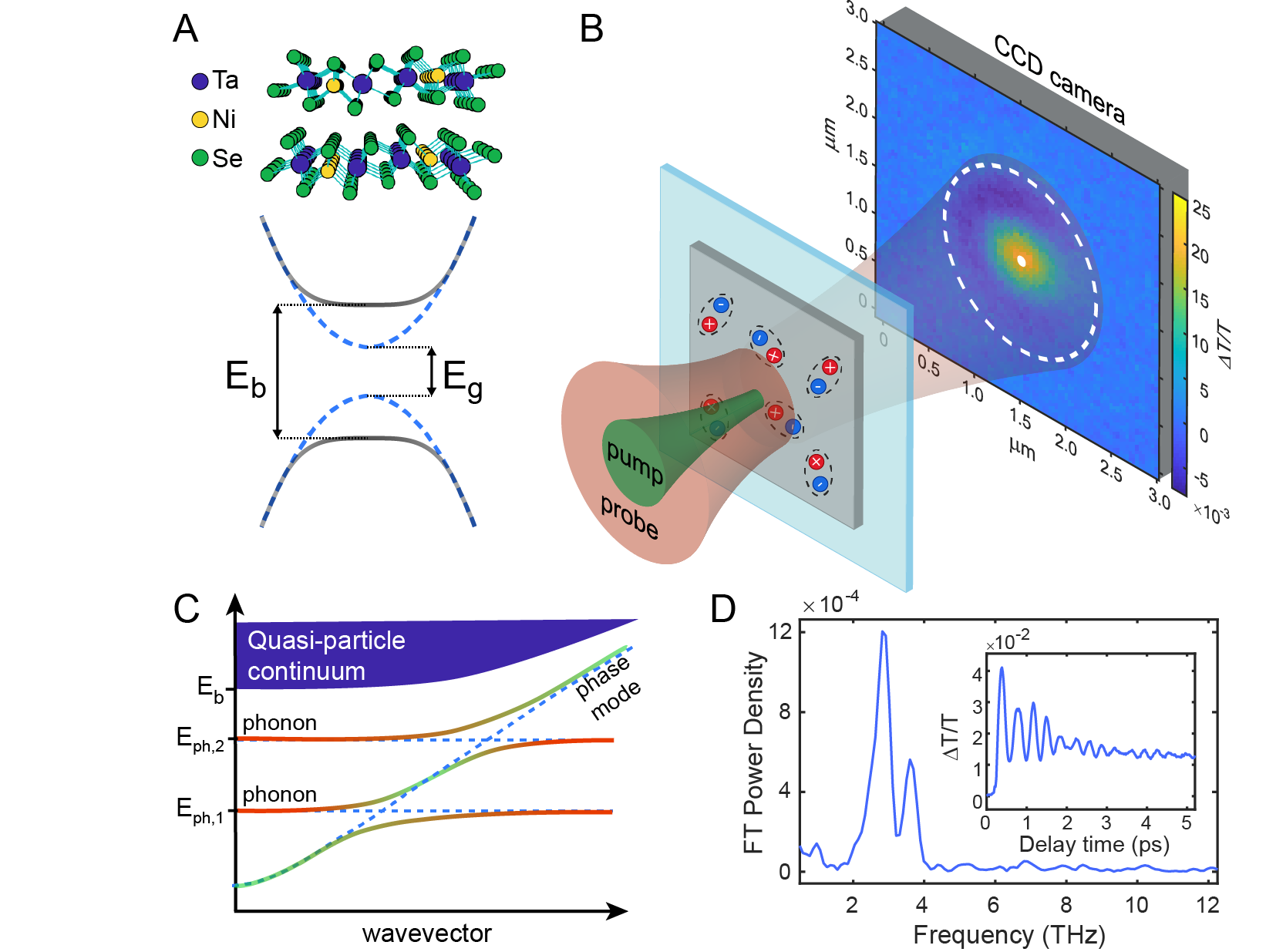}
\caption{Pump-probe microscopy of Ta$_2$NiSe$_5$. (A, top) Crystal structure of Ta$_2$NiSe$_5$. The alternating chains of Ta and Ni atoms confer a quasi-1D nature to the material and independently host conduction and valence band states [\protect\citen{Seki2014PRB}].   (A, bottom) Schematic electronic band structure for a prototypical excitonic insulator. Above the critical temperature the material is close to the semiconductor-semimetal transition (portrayed here as a semiconductor of energy gap E$_g$). Below the critical temperature, the exciton binding energy (E$_b$) exceeds the single-particle gap (E$_g$), resulting in a macroscopic coherent state. (B) Schematic of the measurement setup at the location of the sample with the photoinduced transmissivity change ($\Delta$T/T) signal. The white dot and dashed circle are examples of pixel regions over which we average the signal in our analysis.  (C) Schematic diagram of a possible low-energy excitation structure for an excitonic insulator in the presence of electron-phonon coupling. The phonon modes of the material hybridize with the phase mode resulting in mixed phonon-phase modes as long as the energy gap in the phase mode dispersion at k = 0 is smaller than the phonon energy E$_{ph,i}$. The phonon and phase content of the modes is represented as a color gradient from green (pure phase mode) to red (pure phonon mode). (D) Photoinduced transmissivity change ($\Delta$T/T) as a function of the pump-probe delay time collected at the center of the pump region (inset) and its Fourier Transform power density.}

\label{fig1}
\end{figure*}

We probe the spatio-temporal dynamics of Ta$_2$NiSe$_5$ following photoexcitation using a recently developed femtosecond optical pump-probe microscopy technique [\citen{hartland2010ultrafast,schnedermann2019ultrafast,sung2020long}], which provides sub 10 fs time-resolution and 10 nm spatial precision (for a characterization of the sample with standard, non-spatially resolved pump-probe spectroscopy, see Supplementary Materials Sec.~S1). In these measurements (see Fig.~\ref{fig1}B) we perturb a small area of the sample using a diffraction limited ($\simeq$400 nm full-width half-maximum) optical pulse of $\simeq$12 fs duration. We measure the resulting change in transmission using a wide-field (15 $\mu$m full-width half-maximum), $\simeq$10 fs probe pulse, which is projected onto an EMCCD camera. The large probe allows us to study both the area directly excited, and the surrounding material, imaging how excitations propagate in space and time as the presence of quasiparticles or oscillatory modes in the system modulate the probe transmission. Fig.~\ref{fig1}B displays a schematic of the microscopy setup at the position of the sample and an example of a two-dimensional plot of the photoinduced transmissivity change ($\Delta$T/T) as measured 300 fs after the arrival of the pump pulse.  $\Delta$T/T is obtained from (T$_{\text{pump on}}$-T$_{\text{pump off}}$)/(T$_{\text{pump off}}$), where  T$_{\text{pump on}}$ and T$_{\text{pump off}}$ indicate the transmission measured in the presence and in the absence of the pump pulse respectively. Crucially, as we describe below, we primarily focus our analysis on the regions of the material where we do not directly inject hot quasiparticles.

The kinetic recorded at the center of the excitation spot as a function of the pump-probe delay time is shown in the inset of Fig.~\ref{fig1}D and is analogous to those obtained in previous works using non-spatially resolved pump-probe spectroscopy [\citen{werdehausen2018coherent,mor2018inhibition}]. The signal is characterized by an exponentially decaying electronic component superimposed with a strong oscillatory signal. In Fig.~\ref{fig1}D we additionally report the Fourier Transform (FT) of this oscillatory component. The main recognizable features are the peaks at 1, 2.9 and 3.6 THz, which have been commonly associated with phonon modes in \TNS{}  [\citen{werdehausen2018coherent,mor2018inhibition}].

\begin{figure*}[!t]
    \centering

\includegraphics[]{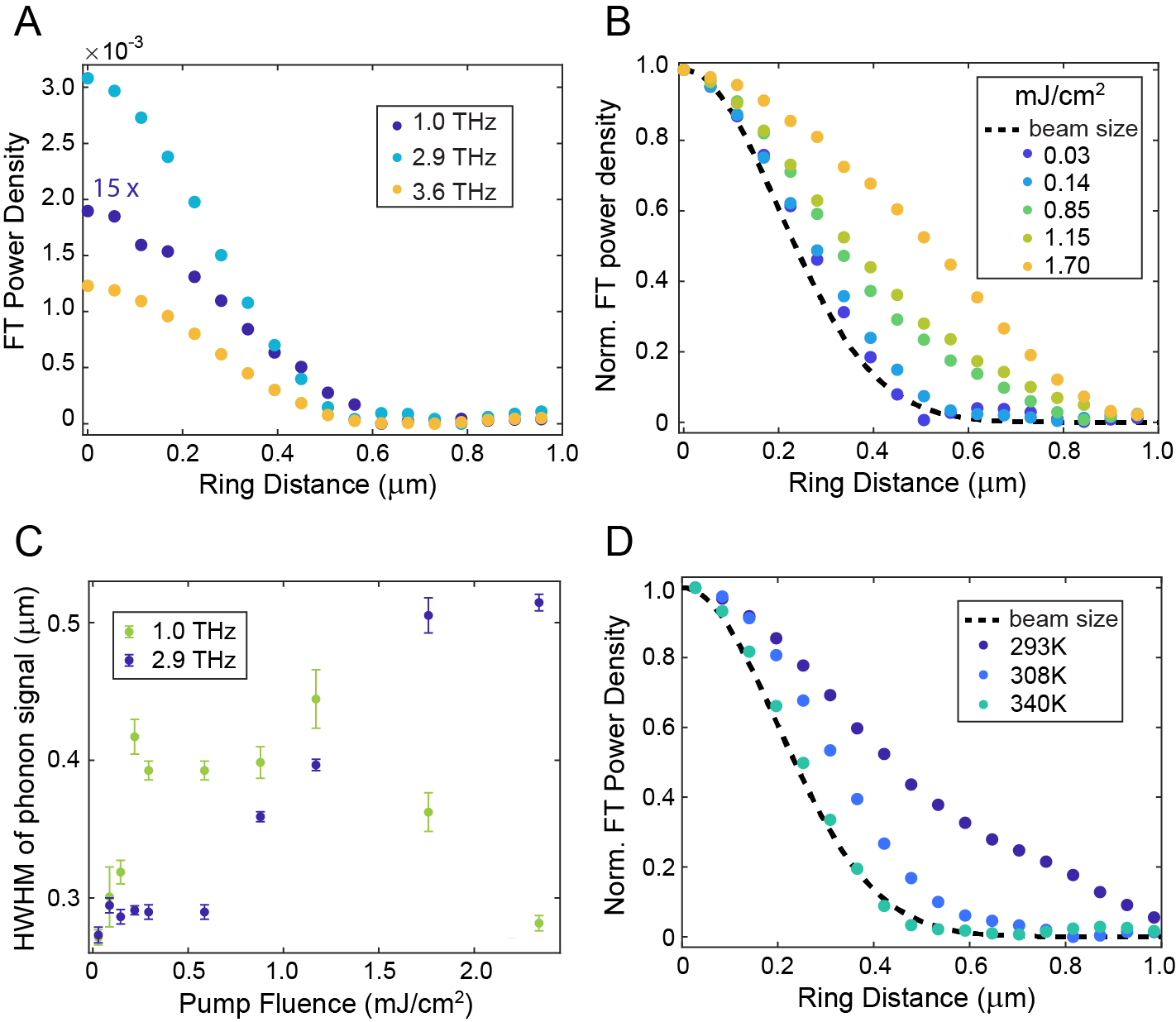}
\caption{Temperature and fluence dependent propagation. (A) Fourier Transform (FT) power density as a function of the distance of the ring over which we average the signal from the center of the pump region. This data was collected using a pump fluence of 0.3~mJ/cm$^2$. The data set for the 1 THz mode is magnified by a factor of 15 to improve its visibility. (B,D) Spatial dependence of the normalized FT power density for the 2.9 THz mode as a function of pump fluence (measured at room temperature)  (B) and of temperature (each measured at a comparable fluence of $\simeq$ 1.2~mJ/cm$^2$) (D). The dashed line indicates the profile of the pump pulse. (C) Half-width half-maximum (HWHM) of the phonon spatial extension as a function of the pump fluence for the 1 THz and 2.9 THz modes.}

\label{fig2}
\end{figure*}

The same FT analysis is now applied to the signal in areas away from the pump position, where no excitations are generated directly by the pump pulse. In this case, to focus on coherent propagation out of the photoexcited region and to maximize the signal-to-noise ratio, we average the $\Delta$T/T signal over rings of pixels equidistant from the center (see Supplementary Materials Sec.~S2 for details on the nearly isotropic nature of the signal). Through this procedure, we determine the spatial decay of the oscillatory modes by plotting the FT power density as a function of the ring distance, as shown in Fig.~\ref{fig2}A for the 1, 2.9, and 3.6 THz modes. The resulting Gaussian shape is a convolution of the profile of the pump pulse (shown as a reference in Fig.~\ref{fig2}B-D) and of the spatial propagation of the excited modes.

In Fig.~\ref{fig2}B, we show the dependence of the FT power density spatial profile for the 2.9 THz mode on the pump fluence. From a Gaussian fit of these curves we extract the half width at half maximum (HWHM), which is plotted in Fig.~\ref{fig2}C. We observe a highly nonlinear dependence of HWHM against the field strength, as the spatial extent of the FT grows after $\simeq$ 0.55 mJ/cm$^2$ and reaches twice its original size, before finally saturating when we approach the material breaking point. This result suggests that at low fluences the signal is dominated by the profile of the pump pulse and the propagation of the mode outside the pumped region becomes visible only at high fluences. Interestingly, while an analogous behavior is observed for the 3.6 THz mode (see Supplementary Materials Sec.~S3), a rather different fluence dependence emerges for the 1 THz mode (Fig.~\ref{fig2}C). This result hints at different microscopic mechanisms underlying the excitation of the oscillations, indicating nonlinear effects may be required to excite the higher frequency modes.

We next compare (Fig.~\ref{fig2}D) the spatial distribution of the FT power density of the 2.9 THz mode above and below T$_c$ (see the Supplementary Materials Sec.~S3 for analogous behavior in the 1 and 3.6 THz modes).  While the data is normalized for ease of comparison, the fluence, the overall $\Delta$T/T signal, and the absolute FT power density at the center of the pumped region are comparable for all these measurements. As T$_c$ is approached from below, the FT profile narrows. Once the temperature surpasses T$_c$, the FT profile has shrunk to the size of the pump spot size, suggesting that above T$_c$ the coherent propagation of the mode breaks down.

\begin{figure}[tbh]
    \centering

\includegraphics[width=50mm]{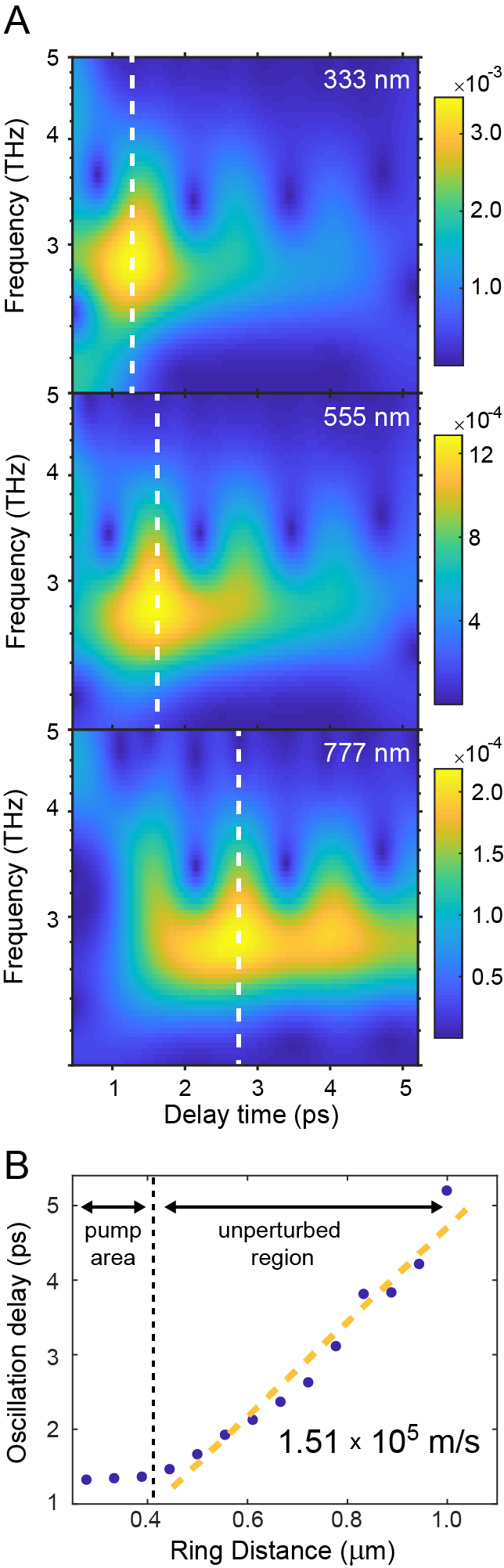}
\caption{Temporal characterization of the oscillatory modes' propagation. (A) Continuous wavelet transform (CWT) of the oscillatory component of the $\Delta$T/T signal at three different distances (333, 555, and 777 nm) from the center of the pump region and measured at a fluence of $\simeq$ 1.6 mJ/cm$^2$. The color map represents the magnitude of the CWT. The white dotted lines approximately mark the location of the strongest phonon oscillations. The periodic oscillations in the CWT are clearly resolved and their possible origins are discussed in the main text. (B) Time delay of the region of strong oscillations as a function of the ring distance. The orange dashed line represents the linear fit of the data in the region outside the pump area and has an inverse slope 1.51$~\times$~10$^5$  m/s.}

\label{fig3}
\end{figure}

Returning to room temperature data, we further analyze the time traces for the different rings by performing continuous wavelet transforms (CWT) (see Supplementary Materials Sec.~S4 for more details). This procedure allows us to determine the amplitude of the oscillatory components as a function of time and distance from the photoexcited area. In Fig.~\ref{fig3}A we show the results of this analysis for a few rings. It can be seen that the strongest amplitude for the 2.9 THz mode (the only one that is clearly traceable within the signal to noise limit) progressively shifts toward later times as we move farther away from the pump region. The position of this region in time is represented in Fig.~\ref{fig3}B for all the rings. By performing a linear fit of this curve, focusing on ring radii residing outside of the directly pumped region, we extract a propagation velocity of 1.51 $\pm$ 0.11 x 10$^5$ m/s. This means that below T$_c$ the 2.9 THz mode propagates coherently, at a velocity characteristic of electronic excitations, for distances of up to 1 $\mu$m. 

\section{Discussion}

We now discuss possible origins of the anomalous spatial propagation at room temperature. First, we note that the measured propagation velocity is orders of magnitude larger than typical values both for optical phonons, which are commonly characterized by rather flat energy dispersions, and acoustic phonons, whose velocity is typically below 10$^3$ m/s~[\citen{gorelik2012dispersion}]. This suggests that the propagating signal does not have a purely phononic origin. Another potential explanation of the observed behavior is the excitation of rapidly propagating bulk or surface phonon-polaritons [\citen{dai2014tunable,basov2016polaritons}]. However, as a result of the crystal inversion symmetry, the bulk phonon modes are Raman and not IR active, and therefore cannot participate in the formation of phonon-polariton modes. In addition, for this type of excitation we would expect to measure higher propagation velocities ($>$5$\times 10^6$~m/s) due to the light-like nature of these modes (see Supplementary Materials Section~S8). Finally, the disappearance of the propagation above T$_C$ is unlikely to be explained by this scenario, as the observed phonon modes characterize both the low and the high temperature phase of this material, which several recent works suggest being an insulator even above T$_C$ [\citen{sugimoto2018strong,andrich2020ultrafast}]. We next consider the possibility of the propagation of quasiparticles (QPs) emanating from the excitation spot with velocity $v_k=(\partial \epsilon _k/\partial k)$ of the order of 10$^5$ m/s, which can excite phonons in their wake. However, at room temperature, the carrier relaxation time of QPs is usually only of the order of a few tens of femtoseconds, corresponding to a mean free path of a few nanometers [\citen{gall2016electron,jablonski1984inelastic}]. In addition, the emission of phonons by the QPs would generate oscillations that are not in phase as we average rings around the photoexcited region. In our experiments coherent behavior is observed on a much wider spatial range and on timescales extending to a few picoseconds. We contend that more coherent electronic oscillations (such as the case of plasmons) are unlikely to be excited in this experiment as the material is an insulator at equilibrium. Even after photoexcitation we therefore do not expect the probed region to host the required carrier population to support these modes.
An exciton-polariton, formed by uncondensed excitons at the conduction-band edge, is a final potential candidate that could explain the observed behavior without requiring the existence of any exotic ground state in the system. Yet, long-range and long-lived (several picosecond) excitations such as the ones that we detect require significant coupling between the light and matter modes and significant confinement. Considering that for comparable van der Waals materials the estimated Q factors for a 60 nm flake are $\leq$5~[\citen{basov2016polaritons}], we would not expect to observe modes with lifetimes longer than $\sim$10 fs in our system (see Supplementary Materials Section~S8). Very strong light-matter coupling (resulting in a Rabi splitting of the excitonic line larger than 100 meV) also would be required to achieve experimentally relevant velocities, yet no hints of this are observed in equilibrium measurements~[\citen{larkin2017}]. Finally, we are unaware of a mechanism by which these modes could be coherently excited with our pumping scheme, where the injection of quasiparticles with energies much larger than the bandgap will relax to the bandedge through successive stochastic relaxation processes. We therefore believe that these considerations make the exciton-polariton scenario rather remote, but analogous measurements in the mid-IR would be useful to more conclusively address this scenario.

One remaining possibility for the anomalous propagation is the coupling between the phonon modes and the EI’s collective phase mode. A phase mode is characterized by a linear dispersion and group velocities on the electronic scale. If electron-phonon coupling is present, the phase and the phonon modes can hybridize resulting in excitations of mixed electronic and phononic nature which can propagate at velocities close to that of the pure phase mode (see Fig.~\ref{fig1}D). Indeed, the presence of long-range exciton-phonon complexes supporting acoustic-like low energy excitations is consistent with recent measurements of the optical response functions of \TNS{}[\citen{larkin2017,larkin2018}]. We expect that the lifetime of the phase mode can be significantly longer than that of QPs, as is supported by a theoretical calculation of the effects of the possible disorder on the QPs and the phase mode (see the Supplementary Materials Sec.~S9 and [\citen{BenjiPRB2020}]).  While the overall relaxation rate of the phonon-phase mode might not differ drastically from that of the phonons, the significantly higher speed can result in long propagation lengths ($\simeq \mu$m) for these excitations. The hybrid modes can then carry a phononic signature ballistically over long distances.  This coupled phonon-phase mode scenario is also consistent with the suppression of propagation above T$_c$, where the collective phase mode should disappear. We note that a hybrid phonon-phase mode with an electronic-like group velocity is realized as long as the gap of the phase mode is comparable or smaller than the phonon energy. This occurs regardless of the origin of the gap, even where additional electronic terms directly breaking the symmetry [\citen{mazza2019nature,watson2019band}] also contribute to the formation of the gap.

Returning to Fig.~\ref{fig2}C, we can tentatively interpret the different behavior of the two oscillatory modes as the result of a different coupling process between these phonons and the phase mode (see Supplementary Materials Sec.~S6). As the hybridization between the two modes occurs at larger wavevector with increasing phonon frequency, we expect that a higher order process, like Raman scattering involving multiple phonon modes,  could be required to excite the more energetic modes at wavevectors that are not provided by the focused laser beam (see Supplementary Materials Sec.~S5). Intriguingly, the saturation and drop in the 1 THz extension at high fluences could be the signature of the enhancement of the gap in the phase mode dispersion. If raised above the phonon energy, the phase mode would indeed not contribute to the formation of the hybrid modes. Further studies that go beyond the scope of this work are required to address these hypotheses.

To support our interpretations, we develop a theoretical model within a two band approximation. We start from a typical Hubbard-type Hamiltonian with additional terms accounting for the electron--phonon coupling and the phonon energy, 

\begin{equation}
H=H_{\text{kin}} + H_{\text{int}} + H_{\text{el-ph}} + H_{\text{ph}}
\end{equation}
 where $H_{\text{kin}}$ represents the electronic bands, $H_{\text{int}}$ is the electron--electron interaction, $H_{\text{el-ph}}$ is the electron--phonon coupling, and $H_{\text{ph}}$ is the phonon Hamiltonian. This approach is analogous to that used in previous theoretical works on Ta$_2$NiSe$_5$ [\citen{Murakami2017PRL,Seki2014PRB,sugimoto2018strong}], and like there we use a form of $H_{\text{el-ph}}$ that explicitly breaks the symmetry of $H$ (see Supplementary Materials Sec.~S7 and [\citen{YutaPRB2020}] for more details, calculations at various temperatures, and for an analysis of the opposite case with no explicit symmetry breaking). The microscopic parameters corresponding to the band structure and the electron--electron interaction were estimated by fitting previous experimental ARPES results [\citen{Seki2014PRB}] (see Supplemental Materials Sec.~S7B). In Fig.~\ref{fig4}A, the calculated linear response function shows the dispersion of the massless phase mode in the absence of electron--phonon coupling. Using this dispersion, the estimated group velocity is $v_{\text{PM}}$= 1.0 x 10$^5$ m/s for this mode, which is of the same order of magnitude of the velocity observed in the experiments. In Fig.~\ref{fig4}B, we show the result in the presence of the electron--phonon coupling term where the phase mode becomes massive and a hybridization between the phase and phonon mode occurs.

In order to determine whether the phonon oscillations induced by the excitation can propagate at velocities of the order of $v_{\text{PM}}$ (an electronic-like velocity), we perform a real-space time-dependent mean-field simulation (see Supplementary Materials  Sec.~7A). Here, we use a larger phonon frequency (indicated by the red, dashed line in Fig.~\ref{fig4}B, D-F) than those observed in the experimental data and we correspondingly consider a smaller excitation area. These assumptions do not affect the conclusions that we can draw from these calculations and they are adopted only for the sake of simplifying the numerical simulation. In Fig.~\ref{fig4}C, we show the propagation of the phonon displacement after a spatially confined excitation. The signal propagates with a velocity of about 0.6$\times$10$^5$ m/s. We can visualize the time evolution of the phonon displacement at different distances from the excitation by performing a spatially windowed Fourier Transform, analogous to a CWT analysis. The results are shown in Fig.~\ref{fig4}D-F, where the orange dashed line indicates the frequency associated with the gap in the phase mode dispersion for the parameters used in our simulations. These plots illustrate the spreading of the hybrid mode at the phonon frequency with a velocity comparable to the phase mode velocity and further support the above scenario that mixing between the phase mode and the phonon mode leads to the fast propagation of the phonon oscillations.

We note that in Fig.~\ref{fig4}D-F, oscillations in the phonon displacement as a function of time contain information about the gap of the phase mode at $k\simeq 0$  (see the Supplementary Materials Sec.~S7B for details). While the data in Fig.~\ref{fig3}A shows analogous oscillations, we presently cannot determine if their origin is connected to the gapped phase mode or to a more trivial frequency beating between the 2.9 and 3.6 THz modes.

\section{Conclusions}


Using ultrafast pump-probe microscopy, we image the spatial dynamics of oscillatory modes in the EI candidate \TNS{}. Below 328 K, we observe the anomalous propagation of these modes, which remain coherent up to $\simeq$ 1 $\mu$m, at a velocity of $1.5\times10^5$ m/s. After carefully considering various scenarios, we attribute this propagation to the hybridization of dispersionless phonon modes with the dispersive, low-lying phase mode of an excitonic condensate, which are excited through the abrupt change in the quasi-particle density in the excitation region [\citen{cheng1991mechanism}]. Further studies are required to establish the precise processes by which these collective modes are excited. Our results and interpretations have several implications. Firstly, the experimental observation of the signature of a phase mode provides us with important insight into the nature of the excitonic order in \TNS{}. Namely, if the gap of the phase mode was much larger than the phonon frequency, the mixing between the modes would be suppressed. Hence, the effects of the 
electron--phonon coupling and of possible electronic terms that explicitly break the continuous symmetry [\citen{mazza2019nature,watson2019band,zenker2014fate,Murakami2017PRL}] should be weak, indicating that the ordered phase in Ta$_2$NiSe$_5$ is an excitonic insulator phase primarily driven by interband Coulomb interactions (see Supplementary Materials Sec.~S7). Second, a notable experimental observation is the small anisotropy in the mode propagation. This may seem counterintuitive since Ta$_2$NiSe$_5$ is a quasi-one-dimensional system and previous DFT calculations found that the bandwidth of the valence band along the direction perpendicular to the atomic chains should be almost five times smaller than in the direction along the chains [\citen{kaneko2013orthorhombic}]. However, we show theoretically that even in the presence of anisotropic hopping between the lattice sites, when the system approaches the BCS-BEC crossover regime, the anisotropy of the phase-mode velocity becomes small (see Supplementary Materials Sec.~S7B3). Hence, the small experimentally observed anisotropy indicates that the system is close to the BCS-BEC crossover regime, as suggested in previous works [\citen{lu2017zero}]. 
Finally, our results showcase the possibility of using spatially–resolved, femtosecond pump-probe measurements to gain crucial insights into the properties of condensed matter systems manifesting emergent many-body phenomena. By exciting and detecting coherent collective modes and by studying their propagation properties, important information can be gathered on the microscopic origin of these effects. Femtosecond pulses could thus provide a means to control and read out the properties of these systems in future quantum information applications. \\ 
\newpage

\section*{Materials and Methods}

\noindent \textbf{Sample preparation}

\noindent Ta$_2$NiSe$_5$ crystals were grown using the procedure outlined in reference [\citen{pal2020destabilizing}].\\

\noindent  To prepare samples for optical measurements, flakes were exfoliated using gold-assisted exfoliation [\cite{desai2016gold}] onto glass coverslips and subsequently encapsulated with a second glass slide while inside a nitrogen glovebox.\\ \\

\noindent \textbf{Pump-probe microscopy}
A Yb:KGW laser system (Pharos, Light Conversion), delivers 1030 nm light with pulses of 200 fs duration and 30 $\mu$J power at a repetition rate of 200 kHz. The output beam is split to seed two broadband white-light continuum generation stages, used to create pump and probe pulses. The broadband probe beam is generated using a 3 mm YAG crystal, and a fused-silica prism is then used to select the spectral range of interest (from 650 nm to 950 nm). The white-light continuum employed for pump pulses is instead generated using a 3 mm sapphire crystal, delivering pulses covering the wavelength range from 500 nm to 650 nm (a long wavelength cut off is achieved using a short-pass filter (FESH650, Thorlabs)). The pulses are temporally compressed using chirped mirrors and fused silica wedges, to achieve <12 fs pulses. In particular, two sets of Layertec (109811) mirrors are used on the pump path to also compensate for the glass elements contained in the objective; Venteon (DCM9) (Venteon) mirrors are instead used in the probe path. The pulse length is verified using a frequency resolved optical gating technique [\citen{trebino1997measuring}]. A mechanical chopper (Thorlabs, MC2000B) is used to modulate the pump pulse stream at a frequency of 30 Hz, and a 40 $\mu$m pinhole is used to spatially clean the laser mode. The pump beam is finally expanded to achieve proper filling of the back aperture of an oil-immersion, 1.1 NA objective, which finally focuses the pulses onto the sampleto a spot size of $\sim$400 nm full-width-half-maximum (FWHM). Reflected components of the pump are removed from the collection path using a longpass filter. The probe pulses are mechanically delayed with respect to the pump excitation using a closed-loop piezo translation stage (P-625.1CL, Physik Instrumente), and then focused onto the sample to a size of $\sim$15 $\mu$m FWHM using a spherical mirror. The transmitted probe pulse is collected by the objective, sent through a 10 nm bandpass filter (Thorlabs) to select the wavelength range of interest, and is finally imaged onto an EMCCD camera (Rolera Thunder, QImaging).  For further details on the experimental apparatus, see reference \citen{schnedermann2019ultrafast}. 

\newpage

\bibliographystyle{unsrt}

\bibliography{EIMain.bbl}

\begin{thebibliography}{10}

\bibitem{jerome1967excitonic}
D.~J{\'e}rome, T.~M. Rice, and W.~Kohn.
\newblock Excitonic insulator.
\newblock {\em Physical Review}, 158(2):462, 1967.

\bibitem{halperin1968possible}
B.~I. Halperin and T.~M. Rice.
\newblock Possible anomalies at a semimetal-semiconductor transistion.
\newblock {\em Reviews of Modern Physics}, 40(4):755, 1968.

\bibitem{keldysh1968collective}
L.~V. Keldysh and A.~N. Kozlov.
\newblock Collective properties of excitons in semiconductors.
\newblock {\em Sov. Phys. JETP}, 27(3):521, 1968.

\bibitem{coleman2015introduction}
P.~Coleman.
\newblock {\em Introduction to many-body physics}.
\newblock Cambridge University Press, 2015.

\bibitem{Hanamura1974}
E.~Hanamura and H.~Haug.
\newblock Will a bose-condensed exciton gas be superfluid?
\newblock {\em Solid State Communications}, 15(9):1567 -- 1570, 1974.

\bibitem{cercellier2007evidence}
H.~Cercellier, C.~Monney, F.~Clerc, C.~Battaglia, L.~Despont, M.~G. Garnier,
  H.~Beck, P.~Aebi, L.~Patthey, H.~Berger, et~al.
\newblock Evidence for an excitonic insulator phase in 1{T-TiSe}$_2$.
\newblock {\em Phys. Rev. Lett.}, 99(14):146403, 2007.

\bibitem{monney2011exciton}
C.~Monney, C.~Battaglia, H.~Cercellier, P.~Aebi, and H.~Beck.
\newblock Exciton condensation driving the periodic lattice distortion of
  {1T-TiSe}$_2$.
\newblock {\em Phys. Rev. Lett.}, 106(10):106404, 2011.

\bibitem{Seki2014PRB}
K.~Seki, Y.~Wakisaka, T.~Kaneko, T.~Toriyama, T.~Konishi, T.~Sudayama, N.~L.
  Saini, M.~Arita, H.~Namatame, M.~Taniguchi, et~al.
\newblock Excitonic bose-einstein condensation in {Ta}$_2${NiSe}$_5$ above room
  temperature.
\newblock {\em Phys. Rev. B}, 90:155116, 2014.

\bibitem{mor2017ultrafast}
S.~Mor, M.~Herzog, D.~Gole{\v{z}}, P.~Werner, M.~Eckstein, N.~Katayama,
  M.~Nohara, H.~Takagi, T.~Mizokawa, C.~Monney, et~al.
\newblock Ultrafast electronic band gap control in an excitonic insulator.
\newblock {\em Phys. Rev. Lett.}, 119(8):086401, 2017.

\bibitem{werdehausen2018coherent}
D.~Werdehausen, T.~Takayama, M.~H{\"o}ppner, G.~Albrecht, A.~W. Rost, Y.~Lu,
  D.~Manske, H.~Takagi, and S.~Kaiser.
\newblock Coherent order parameter oscillations in the ground state of the
  excitonic insulator {Ta}$_2${Ni}{Se}$_5$.
\newblock {\em Sci. Adv.}, 4(3):eaap8652, 2018.

\bibitem{kogar2017signatures}
A.~Kogar, M.~S. Rak, S.~Vig, A.~A. Husain, F.~Flicker, Y.~I. Joe, L.~Venema,
  G.~J. MacDougall, T.~C. Chiang, E.~Fradkin, et~al.
\newblock Signatures of exciton condensation in a transition metal
  dichalcogenide.
\newblock {\em Science}, 358(6368):1314--1317, 2017.

\bibitem{wakisaka2009excitonic}
Y.~Wakisaka, T.~Sudayama, K.~Takubo, T.~Mizokawa, M.~Arita, H.~Namatame,
  M.~Taniguchi, N.~Katayama, M.~Nohara, and H.~Takagi.
\newblock Excitonic insulator state in {Ta}$_2${Ni}{Se}$_5$ probed by
  photoemission spectroscopy.
\newblock {\em Phys. Rev. Lett.}, 103(2):026402, 2009.

\bibitem{kaneko2013orthorhombic}
T.~Kaneko, T.~Toriyama, T.~Konishi, and Y.~Ohta.
\newblock Orthorhombic-to-monoclinic phase transition of {Ta}$_2${Ni}{Se}$_5$
  induced by the {B}ose-{E}instein condensation of excitons.
\newblock {\em Phys. Rev. B}, 87(3):035121, 2013.

\bibitem{lu2017zero}
Y.~F. Lu, H.~Kono, T.~I. Larkin, A.~W. Rost, T.~Takayama, A.~V. Boris,
  B.~Keimer, and H.~Takagi.
\newblock Zero-gap semiconductor to excitonic insulator transition in
  {Ta}$_2${Ni}{Se}$_5$.
\newblock {\em Nat. Comm.}, 8:14408, 2017.

\bibitem{mor2018inhibition}
S.~Mor, M.~Herzog, J.~Noack, N.~Katayama, M.~Nohara, H.~Takagi, A.~Trunschke,
  T.~Mizokawa, C.~Monney, and J.~St{\"a}hler.
\newblock Inhibition of the photoinduced structural phase transition in the
  excitonic insulator {Ta}$_2${Ni}{Se}$_5$.
\newblock {\em Phys. Rev. B}, 97(11):115154, 2018.

\bibitem{chen2020doping}
L~Chen, TT~Han, C~Cai, ZG~Wang, YD~Wang, ZM~Xin, and Y~Zhang.
\newblock Doping-controlled transition from excitonic insulator to semimetal in
  {Ta}$_2${Ni}{Se}$_5$.
\newblock {\em Phys. Rev. B}, 102(16):161116, 2020.

\bibitem{suzuki2020detecting}
Takeshi Suzuki, Yasushi Shinohara, Yangfan Lu, Mari Watanabe, Jiadi Xu,
  Kenichi~L Ishikawa, Hide Takagi, Minoru Nohara, Naoyuki Katayama, Hiroshi
  Sawa, et~al.
\newblock Detecting electron-phonon couplings during photo-induced phase
  transition.
\newblock {\em arXiv preprint arXiv:2002.10037}, 2020.

\bibitem{tang2020non}
Tianwei Tang, Hongyuan Wang, Shaofeng Duan, Yuanyuan Yang, Chaozhi Huang,
  Yanfeng Guo, Dong Qian, and Wentao Zhang.
\newblock Non-coulomb strong electron-hole binding in {Ta}$_2${Ni}{Se}$_5$
  revealed by time- and angle-resolved photoemission spectroscopy.
\newblock {\em Phys. Rev. B}, 101:235148, Jun 2020.

\bibitem{kim2020direct}
Kwangrae Kim, Hoon Kim, Jonghwan Kim, Changil Kwon, Jun~Sung Kim, and BJ~Kim.
\newblock Direct observation of excitonic instability in {Ta}$_2${Ni}{Se}$_5$.
\newblock {\em arXiv preprint arXiv:2007.08212}, 2020.

\bibitem{volkov2020critical}
Pavel~A Volkov, Mai Ye, Himanshu Lohani, Irena Feldman, Amit Kanigel, Kristjan
  Haule, and Girsh Blumberg.
\newblock Critical charge fluctuations and quantum coherent state in excitonic
  insulator {Ta}$_2${Ni}{Se}$_5$.
\newblock {\em arXiv preprint arXiv:2007.07344}, 2020.

\bibitem{saha2101photo}
Tanusree Saha, Denis Golez, Giovanni De~Ninno, Jernej Mravlje, Yuta Murakami,
  Barbara Ressel, Matija Stupar, and Primoz~Rebernik Ribic.
\newblock Photo-induced phase transition and associated time scales in the
  excitonic insulator {Ta}$_2${Ni}{Se}$_5$.
\newblock {\em arXiv preprint arXiv:2101.03202}, 2021.

\bibitem{pal2020destabilizing}
Sukanya Pal, Shivani Grover, Luminita Harnagea, Prachi Telang, Anupam Singh,
  DVS Muthu, UV~Waghmare, and AK~Sood.
\newblock Destabilizing excitonic insulator phase by pressure tuning of
  exciton-phonon coupling.
\newblock {\em Phys. Rev. Research}, 2(4):043182, 2020.

\bibitem{okazaki2018photo}
K.~Okazaki, Y.~Ogawa, T.~Suzuki, T.~Yamamoto, T.~Someya, S.~Michimae,
  M.~Watanabe, Y.~Lu, M.~Nohara, H.~Takagi, et~al.
\newblock Photo-induced semimetallic states realised in electron-hole coupled
  insulators.
\newblock {\em Nat. Comm.}, 9(1):1--6, 2018.

\bibitem{ning2020signatures}
H~Ning, O~Mehio, M~Buchhold, T~Kurumaji, G~Refael, JG~Checkelsky, and D~Hsieh.
\newblock Signatures of ultrafast reversal of excitonic order in
  {Ta}$_2${Ni}{Se}$_5$.
\newblock {\em Phys. Rev. Lett.}, 125(26):267602, 2020.

\bibitem{andrich2020ultrafast}
Paolo Andrich, Hope~M Bretscher, Prachi Telang, Anupam Singh, Luminita Harnaga,
  Ajay~K Sood, and Akshay Rao.
\newblock Ultrafast melting and recovery of collective order in the excitonic
  insulator {Ta}$_2${Ni}{Se}$_5$.
\newblock {\em arXiv preprint arXiv:2007.03368}, 2020.

\bibitem{kim2020observation}
Min-Jae Kim, Armin Schulz, Tomohiro Takayama, Masahiko Isobe, Hidenori Takagi,
  and Stefan Kaiser.
\newblock Phononic soft mode behavior and a strong electronic background across
  the structural phase transition in the excitonic insulator
  {Ta}$_2${Ni}{Se}$_5$.
\newblock {\em Phys. Rev. Research}, 2(4):042039, 2020.

\bibitem{mazza2019nature}
G.~Mazza, M.~R{\"o}sner, L.~Windg{\"a}tter, S.~Latini, H.~H{\"u}bener, A.~J.
  Millis, A.~Rubio, and A.~Geroges.
\newblock Nature of symmetry breaking at the excitonic insulator transition:
  {Ta}$_2${Ni}{Se}$_5$.
\newblock {\em arXiv preprint arXiv:1911.11835}, 2019.

\bibitem{watson2019band}
M.~D. Watson, I.~Markovi{\'c}, E.~A. Morales, P.~Le~F{\`e}vre, M.~Merz, A.~A.
  Haghighirad, and P.~D.~C. King.
\newblock Band hybridisation at the semimetal-semiconductor transition of
  {Ta}$_2${Ni}{Se}$_5$ enabled by mirror-symmetry breaking.
\newblock {\em arXiv preprint arXiv:1912.01591}, 2019.

\bibitem{zenker2014fate}
B.~Zenker, H.~Fehske, and H.~Beck.
\newblock Fate of the excitonic insulator in the presence of phonons.
\newblock {\em Phys. Rev. B}, 90(19):195118, 2014.

\bibitem{baldini2020spontaneous}
Edoardo Baldini, Alfred Zong, Dongsung Choi, Changmin Lee, Marios~H Michael,
  Lukas Windgaetter, Igor~I Mazin, Simone Latini, Doron Azoury, Baiqing Lv,
  et~al.
\newblock The spontaneous symmetry breaking in {Ta}$_2${Ni}{Se}$_5$ is
  structural in nature.
\newblock {\em arXiv preprint arXiv:2007.02909}, 2020.

\bibitem{hartland2010ultrafast}
G.~V. Hartland.
\newblock Ultrafast studies of single semiconductor and metal nanostructures
  through transient absorption microscopy.
\newblock {\em Chemical Science}, 1(3):303--309, 2010.

\bibitem{schnedermann2019ultrafast}
C.~Schnedermann, J.~Sung, R.~Pandya, S.~D. Verma, R.~Y.~S. Chen, N.~Gauriot,
  H.~M. Bretscher, P.~Kukura, and A.~Rao.
\newblock Ultrafast tracking of exciton and charge carrier transport in
  optoelectronic materials on the nanometer scale.
\newblock {\em JPCL}, 10(21):6727--6733, 2019.

\bibitem{sung2020long}
J.~Sung, C.~Schnedermann, L.~Ni, A.~Sadhanala, R.~Y.~S. Chen, C.~Cho,
  L.~Priest, J.~M. Lim, H.-K. Kim, B.~Monserrat, et~al.
\newblock Long-range ballistic propagation of carriers in methylammonium lead
  iodide perovskite thin films.
\newblock {\em Nat. Phy.}, 16(2):171--176, 2020.

\bibitem{gorelik2012dispersion}
V.~S. Gorelik and N.~S. Vasil’ev.
\newblock Dispersion of optical and acoustic phonons in diamond and germanium
  crystals.
\newblock {\em Inorganic Materials}, 48(5):462--468, 2012.

\bibitem{dai2014tunable}
S.~Dai, Z.~Fei, Q.~Ma, A.~S. Rodin, M.~Wagner, A.~S. McLeod, M.~K. Liu,
  W.~Gannett, W.~Regan, K.~Watanabe, et~al.
\newblock Tunable phonon polaritons in atomically thin van der $w$aals crystals
  of boron nitride.
\newblock {\em Science}, 343(6175):1125--1129, 2014.

\bibitem{basov2016polaritons}
DN~Basov, MM~Fogler, and FJ~Garc{\'\i}a De~Abajo.
\newblock Polaritons in van der waals materials.
\newblock {\em Science}, 354(6309), 2016.

\bibitem{sugimoto2018strong}
K.~Sugimoto, S.~Nishimoto, T.~Kaneko, and Y.~Ohta.
\newblock Strong coupling nature of the excitonic insulator state in
  {Ta}$_2${Ni}{Se}$_5$.
\newblock {\em Phys. Rev. Lett.}, 120(24):247602, 2018.

\bibitem{gall2016electron}
D.~Gall.
\newblock Electron mean free path in elemental metals.
\newblock {\em Journal of Applied Physics}, 119(8):085101, 2016.

\bibitem{jablonski1984inelastic}
A.~Jablonski, P.~Mrozek, G.~Gergely, M.~Menhyard, and A.~Sulyok.
\newblock The inelastic mean free path of electrons in some semiconductor
  compounds and metals.
\newblock {\em Surface and interface analysis}, 6(6):291--294, 1984.

\bibitem{larkin2017}
T.~I. Larkin, A.~N. Yaresko, D.~Pr\"opper, K.~A. Kikoin, Y.~F. Lu, T.~Takayama,
  Y.-L. Mathis, A.~W. Rost, H.~Takagi, B.~Keimer, et~al.
\newblock Giant exciton fano resonance in quasi-one-dimensional
  {Ta}$_2${NiSe}$_5$.
\newblock {\em Phys. Rev. B}, 95:195144, 2017.

\bibitem{larkin2018}
T.~I. Larkin, R.~D. Dawson, M.~H\"oppner, T.~Takayama, M.~Isobe, Y.-L. Mathis,
  H.~Takagi, B.~Keimer, and A.~V. Boris.
\newblock Infrared phonon spectra of quasi-one-dimensional {Ta}$_2${Ni}{Se}$_5$
  and {Ta}$_2${Ni}{S}$_5$.
\newblock {\em Phys. Rev. B}, 98:125113, 2018.

\bibitem{BenjiPRB2020}
B.~Remez and N.~R. Cooper.
\newblock Effects of disorder on the transport of collective modes in an
  excitonic condensate.
\newblock {\em Phys. Rev. B}, 101:235129, 2020.

\bibitem{Murakami2017PRL}
Y.~Murakami, D.~Gole\ifmmode~\check{z}\else \v{z}\fi{}, M.~Eckstein, and
  P.~Werner.
\newblock Photoinduced enhancement of excitonic order.
\newblock {\em Phys. Rev. Lett.}, 119:247601, 2017.

\bibitem{YutaPRB2020}
Y.~Murakami, D.~Gole\ifmmode~\check{z}\else \v{z}\fi{}, T.~Kaneko, A.~Koga,
  A.~J. Millis, and P.~Werner.
\newblock Collective modes in excitonic insulators: Effects of electron-phonon
  coupling and signatures in the optical response.
\newblock {\em Phys. Rev. B}, 101:195118, 2020.

\bibitem{cheng1991mechanism}
T.~K. Cheng, J.~Vidal, H.~J. Zeiger, G.~D. M.~S. Dresselhaus, M.~S.
  Dresselhaus, and E.~P. Ippen.
\newblock Mechanism for displacive excitation of coherent phonons in {Sb},
  {Bi}, {Te}, and {Ti}$_2${O}$_3$.
\newblock {\em Applied Physics Letters}, 59(16):1923--1925, 1991.

\bibitem{desai2016gold}
Sujay~B Desai, Surabhi~R Madhvapathy, Matin Amani, Daisuke Kiriya, Mark
  Hettick, Mahmut Tosun, Yuzhi Zhou, Madan Dubey, Joel~W Ager~III, Daryl
  Chrzan, et~al.
\newblock Gold-mediated exfoliation of ultralarge optoelectronically-perfect
  monolayers.
\newblock {\em Advanced Materials}, 28(21):4053--4058, 2016.

\bibitem{trebino1997measuring}
Rick Trebino, Kenneth~W DeLong, David~N Fittinghoff, John~N Sweetser, Marco~A
  Krumb{\"u}gel, Bruce~A Richman, and Daniel~J Kane.
\newblock Measuring ultrashort laser pulses in the time-frequency domain using
  frequency-resolved optical gating.
\newblock {\em Review of Scientific Instruments}, 68(9):3277--3295, 1997.

\end{thebibliography}

\newpage

\noindent \textbf{Acknowledgements and funding} 

\noindent The calculations were run on the Beo05 cluster at the University of Fribourg. We acknowledge the CECAM workshop “Excitonic insulator: New perspectives in long-range interacting systems” at EPFL Lausanne for insights and discussions and for providing the opportunity to start this collaboration. We also thank A. Boris (Max Planck Institute for Solid State Research) and A. Musser (Cornell University) for helpful discussions. Funding: H.M.B., P.A., and A.R. acknowledge support from the Winton Programme for the Physics of Sustainability, the Engineering and Physical Sciences Research Council, and the European Research Council (ERC) under the European Union’s Horizon 2020 research and innovation programme (grant agreement 758826). N.R.C. acknowledges funding from the Engineering and Physical Sciences Research Council (EPSRC grant no. EP/P034616/1). Y.M. thanks the Japan Society for the Promotion of Science and the Japan Science and Technology Agency for funding through KAKENHI grant no. JP19K23425 and JST CREST grant no. JPMJCR1901. P.W. acknowledges funding from the ERC Consolidator grant no. 724103 and from the Swiss National Science Foundation via NCCR Marvel. A.K.S. thanks the Department of Science and Technology, India for support under Nanomission and Year of Science Professorship. L.H. acknowledges the financial support from the Department of Science and Technology (DST), India [grant no. SR/WOS-A/PM-33/2018 (G)] and IISER Pune for providing the facilities for crystal growth and characterization. B.R. acknowledges support from the Cambridge International Trust and Wolfson College, Cambridge. D.G. is supported by the Slovenian Research Agency (ARRS) under program nos. P1-0044 and J1-2455. N.R.C., A.J.M., and D.G. acknowledge support from the Simons Foundation.

\noindent \textbf{Author Contributions}\\
\noindent HMB and PA performed measurements and analyzed the data. PT, AS, and LH worked on the growth and characterization of the material. PA, AKS and AR conceived the experiments. YM, DG, and BR performed theoretical modelling with input from AJM, PW and NRC.  HMB, PA, and AR wrote the paper with input from all authors. \\

\noindent \textbf{Competing Interests}\\
\noindent The authors declare that they have no competing financial interests. \\

\noindent \textbf{Data availability}\\
\noindent The data underlying all figures in the main text is publicly available at \\ https://doi.org/10.17863/CAM.64033 

\newpage

\begin{figure*}[tbh]
    \centering

\includegraphics[width=130mm]{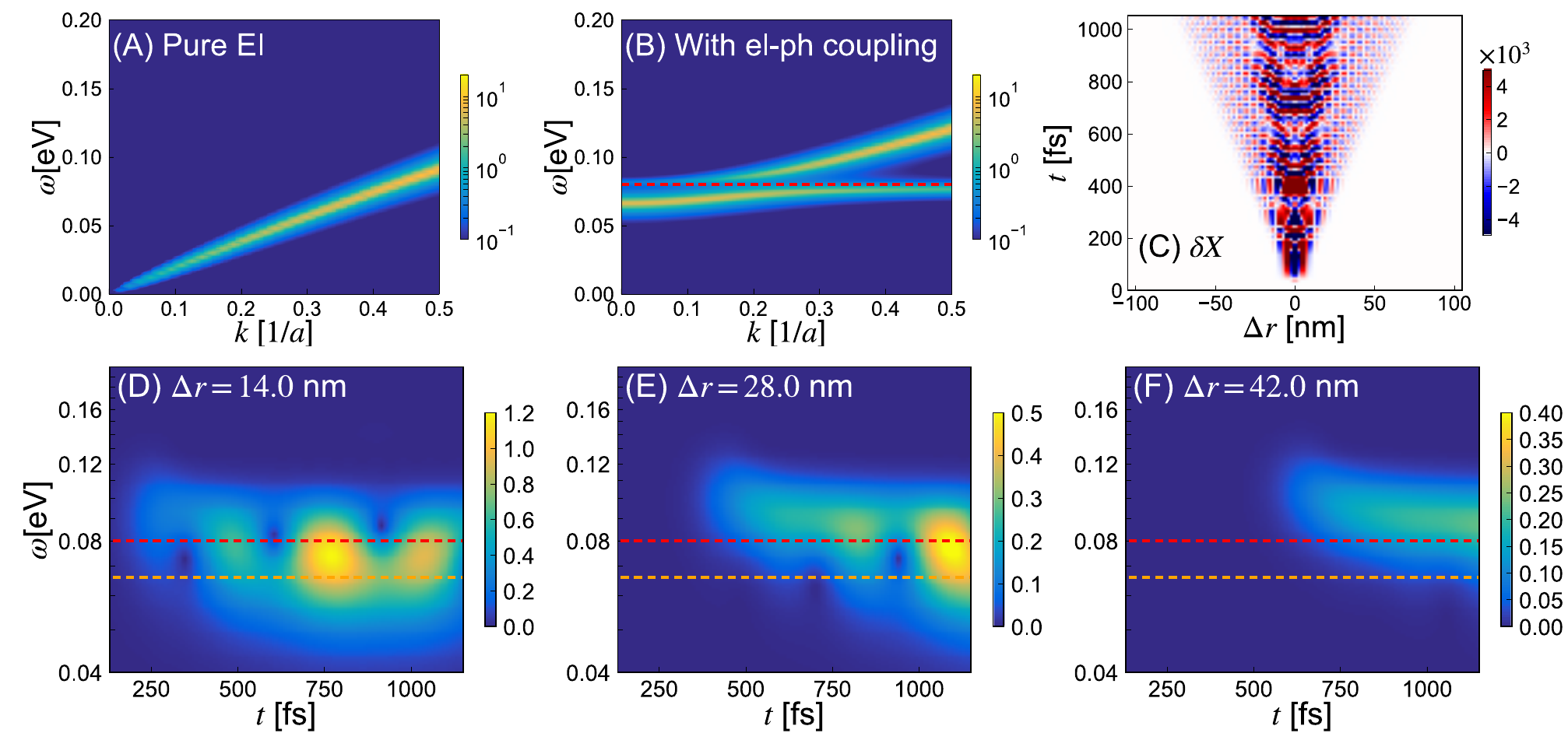}
\caption{Time-dependent Mean-Field Model. (A)(B) Response function of the order parameter for the microscopic two-band model on the one-dimensional lattice with purely excitonic behavior (A) and in the presence of electron--phonon coupling (B). The coupling is small with respect to the Coulomb interaction and explicitly breaks the continuous symmetry of the system. (C) Spatial and temporal evolution of the phonon displacement, $\delta X$ for the case of non-zero electron--phonon coupling following an excitation at the center of the system at time t = 0, calculated within the mean-field description. (D-F) Windowed Fourier Transform of the results in (C) at specified lattice sites, where $\Delta r$ indicates the distance from the center of the excitation area. The red and orange dashed lines indicate the bare phonon frequency used in these calculations and the size of the phase mode gap, respectively. The bandwidth of the conduction and valence bands is set to 1.6 eV and the Coulomb interaction to 0.84 eV. The corresponding band gap is about 0.32 eV.}

\label{fig4}
\end{figure*}



\clearpage

\end{document}